# Laser oscillation in a strongly coupled single quantum dot-nanocavity system

M. Nomura[1*], N. Kumagai[1], S. Iwamoto[1, 2], Y. Ota[1, 2], and Y. Arakawa[1, 2]
[1]Institute for Nano Quantum Information Electronics, The University of Tokyo, 4-6-1 Komaba, Meguro, Tokyo 153-8505, Japan
[2]Institute of Industrial Science, The University of Tokyo, 4-6-1 Komaba, Meguro, Tokyo 153-8505, Japan

Strong coupling of photons and materials[1] in semiconductor nanocavity systems has been investigated because of its potentials in quantum information processing[2] and related applications, and has been testbeds for cavity quantum electrodynamics (QED)[3,4]. Interesting phenomena such as coherent exchange of a single quantum between a single quantum dot and an optical cavity, called vacuum Rabi oscillation[5-9], and highly efficient cavity QED lasers[10-18] have been reported thus far. The coexistence of vacuum Rabi oscillation and laser oscillation appears to be contradictory in nature, because the fragile reversible process may not survive in laser oscillation. However, recently, it has been theoretically predicted that the strong-coupling effect could be sustained in laser oscillation in properly designed semiconductor systems[19]. Nevertheless, the experimental realization of this phenomenon has remained difficult since the first demonstration of the strong-coupling[5, 6], because an extremely high cavity quality factor and strong light-matter coupling are both required for this purpose. Here, we demonstrate the onset of laser oscillation in the strong-coupling regime in a single quantum dot (SQD)-cavity system. A high-quality semiconductor optical nanocavity and strong SQD-field coupling enabled to the onset of lasing while maintaining the fragile coherent exchange of quanta between the SQD and the cavity. In addition to the interesting physical features, this device is seen as a prototype of an ultimate solid state light source with an SQD gain, which operates at ultra-low power, with expected applications in future nanophotonic integrated systems and monolithic quantum information devices.

The confinement of photons in an extremely small volume causes a strong interaction between light and matter. Semiconductor microcavity systems[20] exhibit characteristic physics that can be described by cavity quantum electrodynamics; for example, vacuum Rabi splitting in the strong-coupling regime and laser emission with extremely efficient spontaneous emission coupling in the weak-coupling regime[10-18] have been observed. In the strong-coupling regime, reversible exchange of a single quantum between an SQD and cavity is well-preserved, while irreversible emission dynamics dominate the system in the weak-coupling regime. A recent theoretical study revealed that properly designed SQD-nanocavity systems enable onset of lasing under the strong-coupling condition[19]. In this study, we demonstrate the direct transition from a strong-coupling to a lasing regime in a strongly coupled SQD-nanocavity system. The experimental and theoretical investigations indicated that the laser oscillation began in the strong-coupling regime. The measured photon emission spectra and statistics are in good agreement with theoretical predictions obtained by computing photon emission from a four-level single atom-cavity coupling system on the basis of a quantum master equation model. These results indicate that a majority of the cavity photons (>90%) are provided by the SQD at the threshold.

In general, microcavity systems contain tens or hundreds of QDs per cavity. Therefore, the target SQD is affected by the surrounding QDs, hindering access to the delicate physics of an SQD-cavity system. This deviation in behaviour from an isolated quantum system can be minimized by employing a small cavity in a wafer with an extremely low areal density of QDs. Because of the small mode volume and high cavity quality factor ($Q$), the use of a photonic crystal (PhC) nanocavity[21] is one of the most promising approaches to observe strong-coupling and specifically laser emission in an SQD-nanocavity system.

We used a high-$Q$ PhC nanocavity with a small mode volume and a single, self-assembled indium arsenide (InAs) QD to implement an SQD-cavity system. Three-dimensional photon confinement was obtained by fabricating a PhC slab structure with a nanocavity composed of three missing air holes[22]. This structure confines photons within an extremely small mode volume of $V_{\mathrm{m}} \sim 0.7(\lambda/n)^3$, as shown in the lower right inset of Fig. 1a; the system was simulated using a finite-difference time-domain method. Here, $\lambda$ denotes the wavelength of the cavity mode in vacuum and $n$ = 2.9 is the effective refractive index. The mode volume of the cavity was then calculated to be ~0.02 μm$^3$. The areal density of self-assembled InAs QDs in our semiconductor wafer was ~4 per μm$^2$. Therefore, the average number of QDs in the cavity was only unity. The measured photoluminescence (PL) spectrum at 6 K (Fig. 1b) consisted of a single exciton and cavity mode (estimated $Q$ ~35,000).

The exciton-mode coupling in our system was finely controlled using a temperature-tuning technique, in which an exciton line was scanned through the cavity resonance as shown in Fig. 1c. This technique tuned the relative spectral positions of a target QD and the fundamental cavity mode on the basis of the different temperature dependences of the bandgap and of the refractive index. PL spectra were recorded at an irradiated pump power (defined as the power at the sample surface) of ~3 nW as a function of the temperature. In the temperature tuning measurement, typical phenomena in the strong-coupling regime, such as anti-crossing and energy mixing between the two modes, were observed. The spectra measured in the vicinity of zero detuning of the exciton and cavity mode exhibited an exciton-polariton doublet with approximately identical intensity and linewidth (Fig. 1d). The estimated exciton-mode coupling strength $g$ was 68 μeV. This strength is sufficiently larger than the homogeneous decay rate of the QD exciton, which is generally of the order of 1 μeV, and the photon lifetime of the cavity mode of 38 μeV. Therefore, the system is in the strong-coupling regime.

As observed from the clear doublet feature in the PL spectra, coherent population oscillation between the excitonic and cavity mode occurred in a weak pumping regime in the strongly-coupled system. As the pump power was increased, stimulated emission dominated the dynamics in the system. Figure 2a shows the recorded PL spectra between pump powers of ~25 and 500 nW. The transition from a strong-coupling regime to a lasing regime was clearly observed in this pump power tuning experiment. We found the laser threshold to be ~90 nW by analyzing the light-in versus light-out (L-L) data (Fig. 3a). It is noteworthy that the polariton doublet is still distinct at the threshold pump power. As the pump power was increased, the polariton doublet merged into a single lasing mode located at the bare cavity resonant wavelength and entered a complete lasing regime, where a drastic linewidth narrowing was observed. The asymmetry in the PL spectra in the weak pumping regime was due to the unintentional detuning $\Delta\lambda = \lambda_c - \lambda_e = 0.02$ nm of the excitonic mode $\lambda_e$ and cavity mode $\lambda_c$. We note that our simulation showed that the slight detuning had negligible influence on the main feature. The PL spectra were fitted by two lorentzian functions as shown in Fig. 1d. The analysed L-L plot, linewidths, and peak wavelengths of the two modes are shown in Figs. 3a-3c. The cavity-like mode (red circle) showed a gentle s-shaped L-L plot. Such a soft turn-on lasing is typically observed in microcavity lasers in which spontaneous emission efficiently couples to the lasing mode[11,15,16]. The peak wavelengths of the modes in Fig. 3c show that two distinct polariton branches moved closer to the bare-cavity resonance in the high-pumping regime. The cavity-like mode shows a blue-shift in the high-pump regime due to the carrier-plasma effect, which led to a reduction in the refractive index.

We simulated cavity photon emission in a strongly-coupled SQD-cavity system for better understanding the physics of the system. The model consisted of an incoherently pumped single four-level atom and an optical cavity with $Q$ = 35,000. The quantum dynamics can be described by the quantum master equation in the interaction picture,

$$d\rho_s(t)/dt = -i/\hbar\left[H(t), \rho_s(t)\right] + L\rho_s(t), \tag{1}$$

where $\rho_s(t)$ is a reduced density operator, $H(t)$ denotes an interaction Hamiltonian, and Liouvillian $L$ denotes Markovian processes including spontaneous emission, cavity loss, and incoherent pumping to the SQD. The cavity pumping term was also included in the model to take into account the interfusion of photons from outside the cavity. The pure dephasing process was neglected because the experiments were performed at around 10 K, where the dephasing rate was of the order of 1 GHz[23]. The atom-coupling strength $g$ = 68 μeV and $\Delta\lambda$ = 0.02 nm, which are the parameters of the measured system, were used.

The computed PL spectra and analyzed L-L plots linewidths, and peak wavelengths are shown in Figs. 2b, and 3a-3c, respectively. The computed PL spectra shown in Fig. 2b successfully reproduced the experimental PL spectra shown in Fig. 2a. The general tendency of the analyzed data of the computed PL spectra, shown in Fig. 3d-3f, was also in good agreement with that of the spectra shown in Fig. 3a-3c. In Fig. 3b, the measured linewidth was larger because the spectral convolution of the spectral response function of the detection system and the real spectrum produced the measured spectrum in the experiments. The linewidth increased as the pump power increased due to the pump-induced dephasing. Then, stimulated emission drastically shifts the dynamics in the system from coherent exchange of quantum between the SQD and cavity to irreversible photon emission from the SQD. The linewidth at a high pump regime is narrower than that of a bare cavity. This fact indicates that laser oscillation occurs in the system.

The quantum-statistical characteristics of the photon emission from the system are an important parameter for describing the system. We also calculated the mean cavity photon number $N_{\mathrm{ph}}$ for SQD purity of 90% (red line) and $g^2(0)$ in the steady state, which is defined as $g^2(0) = \langle a^\dagger a^\dagger a a \rangle / \langle a^\dagger a \rangle^2$ for various SQD purities (Fig. 4a). Here, $\langle a^\dagger \rangle$ is the expectation value of the cavity photon creation operator in the steady state. The computed pump rate dependence of $g^2(0)$ for each SQD purity was normalized by the corresponding threshold pump rate[24]. We also measured $g^2(\tau)$ under the strong-coupling condition using a Hanbury Brown-Twiss setup[25]. The experimental $g^2(0)$ values were obtained from the measured values of $g^2(0)$ at various pump powers below the threshold at a temporal resolution of ~400 ps of the setup; these experimental values are plotted using purple circles in Fig. 4a. The result demonstrates that the light emitted from the polariton state (below laser threshold, lower panel of Fig. 4b) was manifestly non-classical, exhibiting sub-Poissonian photon statistics $g^2(0) < 1$ . A comparison of the experimental and computed curves of $g^2(0)$ reveals that the system has an extremely high SQD purity (greater than 90%), which by definition implis that more than 90% of the cavity photons are provided by the SQD at the threshold. This fact indicates that the investigated SQD-cavity system was almost an ideal single artificial atom-cavity system. This SQD-cavity system entered the lasing regime (upper panel of Fig. 4b) with a sufficiently high pump power, where $g^2(0) = 1$ was measured in the photon correlation measurement.

The achievement of laser oscillation in the strong-coupling regime is of considerable interest, because the coherent exchange dynamics is extremely fragile. It has been reported that the strong-coupling state is unexpectedly resistant to dephasing and often appears 'in the disguise' of a single peak[26, 27]. Now, we examine whether the coherent exchange dynamics survives at the laser threshold[24] or not. In our incoherent-pumping system, the basic principle of the strong-coupling condition $g > |\Gamma_{cav} - \Gamma_{ex}|/4$, reported by Valle *et al.*, is applied to our four-level system; the principle is described in the following discussion[27]. Here, we use the effective linewidths of the broadened cavity mode $\Gamma_{cav} = \gamma_{cav} - P_{cav}$ due to cavity pump ($P_{cav}$) and the broadened excitonic mode $\Gamma_{ex} = \gamma_{ex} + (\gamma_{12} + \gamma_{34})/2$, where $\gamma_{ex}$ and $\gamma_{cav}$ are the decay rates of an exciton and the cavity mode without pumping. The term $(\gamma_{12} + \gamma_{34})/2$ corresponds to the mechanism of population-induced dephasing, which affects the phases in the lower (|2>) and upper (|3>) levels of the luminescence process; $\gamma_{12}$ and $\gamma_{34}$ are the respective decay rates from |2> to |1> and from |4> to |3>. In our system, the strong-coupling condition of $2\pi g = 68\mu eV > |\Gamma_{cav} - \Gamma_{ex}|/4$ requires the pump rate of the system to be $P_{sys} < 652$ GHz, which is indicated by the light-blue region in Fig. 4a. The strong-coupling condition is fulfilled at the threshold pump rate of 310 GHz, where the mean cavity photon number reaches unity, and the cavity emission exhibits a high degree of the second order coherence ($g^2(0) \sim 1$) (Fig. 4a). The analyses of experimental results by numerical simulations brought us to the conclusion that laser oscillation, while maintaining a coherent exchange of a quantum between the SQD and cavity, occurred at the pump rate of 310 GHz < $P_{sys}$ < 652 GHz. This result indicates the coexistence of vacuum Rabi oscillation and laser oscillation in the cavity in this pumping regime. The occurrence of laser oscillation in the strong-coupling regime in a system composed of a single trapped gas atom and cavity has been reported [28]. This study reports the first experimental realization of laser oscillation in the strong-coupling regime using a solid-state material.

In conclusion, laser oscillation has been demonstrated in a strongly coupled SQD and PhC nanocavity system. The analyses of the experimental results by numerical simulations on the basis of the quantum master equation reveal that the quantum dynamics of the system directly changes from vacuum Rabi oscillation to laser oscillation without entering the weak-coupling regime. In addition, the comparison between the simulation and experimental results also indicates that the observed polariton states and laser emission occurred in a highly pure (>90%) SQD-cavity coupled system. Such a solid-state single emitter-cavity coupling system is a robust and deterministically reliable system, which includes infallibly a single emitter with a fixed emitter-photon coupling strength. These features encourage the experimental pursuit of the characteristic physics of a single emitter-cavity system.

## Acknowledgements

We thank S. Ishida, M. Shirane, S. Ohkouchi, Y. Igarashi and K. Watanabe for their technical support. M. N. thanks T. Nakaoka, A. Tandaechanurat, S. Kako, and K. Aoki for fruitful discussions. This research was supported by the Special Coordination Funds for Promoting Science and Technology and by KAKENHI 20760030, the Ministry of Education, Culture, Sports, Science and Technology, Japan.

Figures

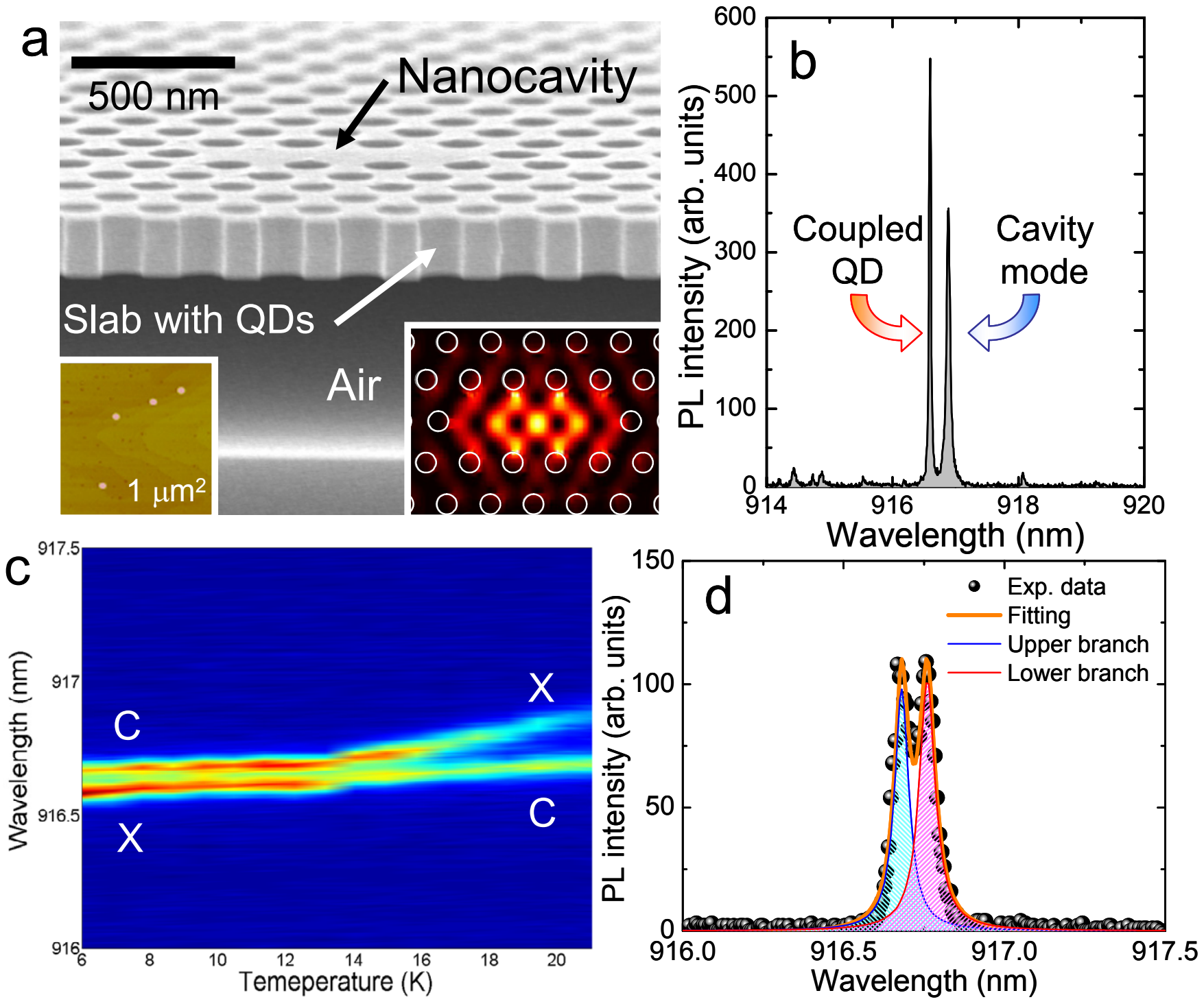

**Figure 1** Photonic crystal structure and optical characteristics. **a**, Scanning electron micrograph of the PhC nanocavity laser. An atomic force microscope image of an equivalent sample without capping (lower left inset). The lower right inset depicts the electric field intensity of the cavity mode, showing that the photons are strongly confined. **b**, PL spectrum of the target exciton and the cavity mode at sufficiently high detuning. **c**, PL spectra recorded at various detunings for a pump power of 3 nW shows vacuum Rabi splitting; x and c denote the exciton and the cavity, respectively. **d**, PL spectrum at zero detuning; clear intensity mixing is observed.

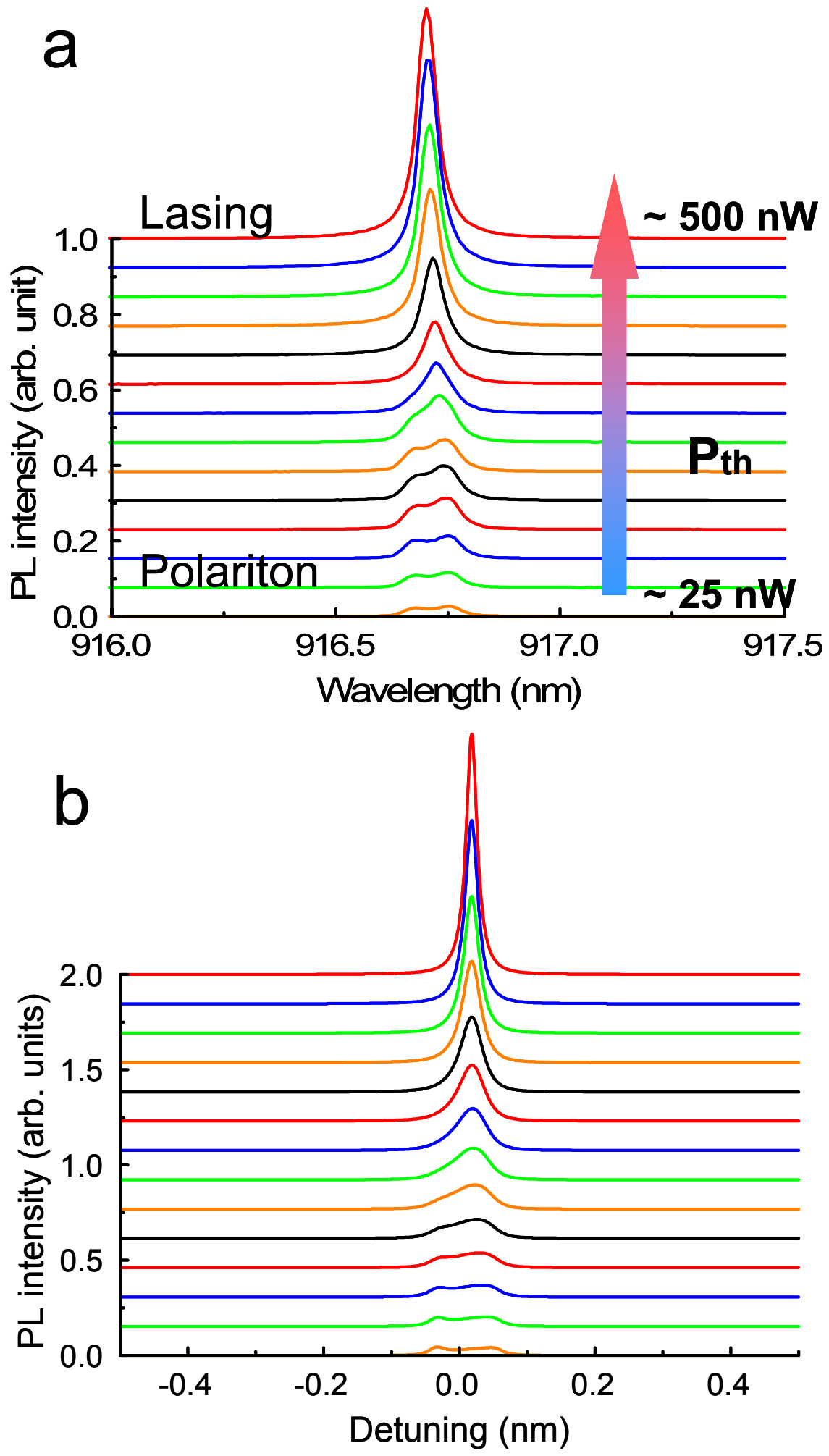

**Figure 2** Experimental and computed PL spectra at various pump powers. **a**, Measured PL spectra recorded between ~25 to 500 nW. Direct transition from the strong-coupling to lasing regime is observed. The wavelength of the lasing mode is identical to the bare cavity resonance. **b**, Computed PL spectra between ~0.1 to $2P_{\mathrm{th}}$: $P_{\mathrm{th}}$ is the threshold pump power. The calculation was carried out using experimentally obtained parameters of $Q$ = 35,000, $g$ = 68 μeV, and $\Delta_{\lambda}$ = 0.02 nm.

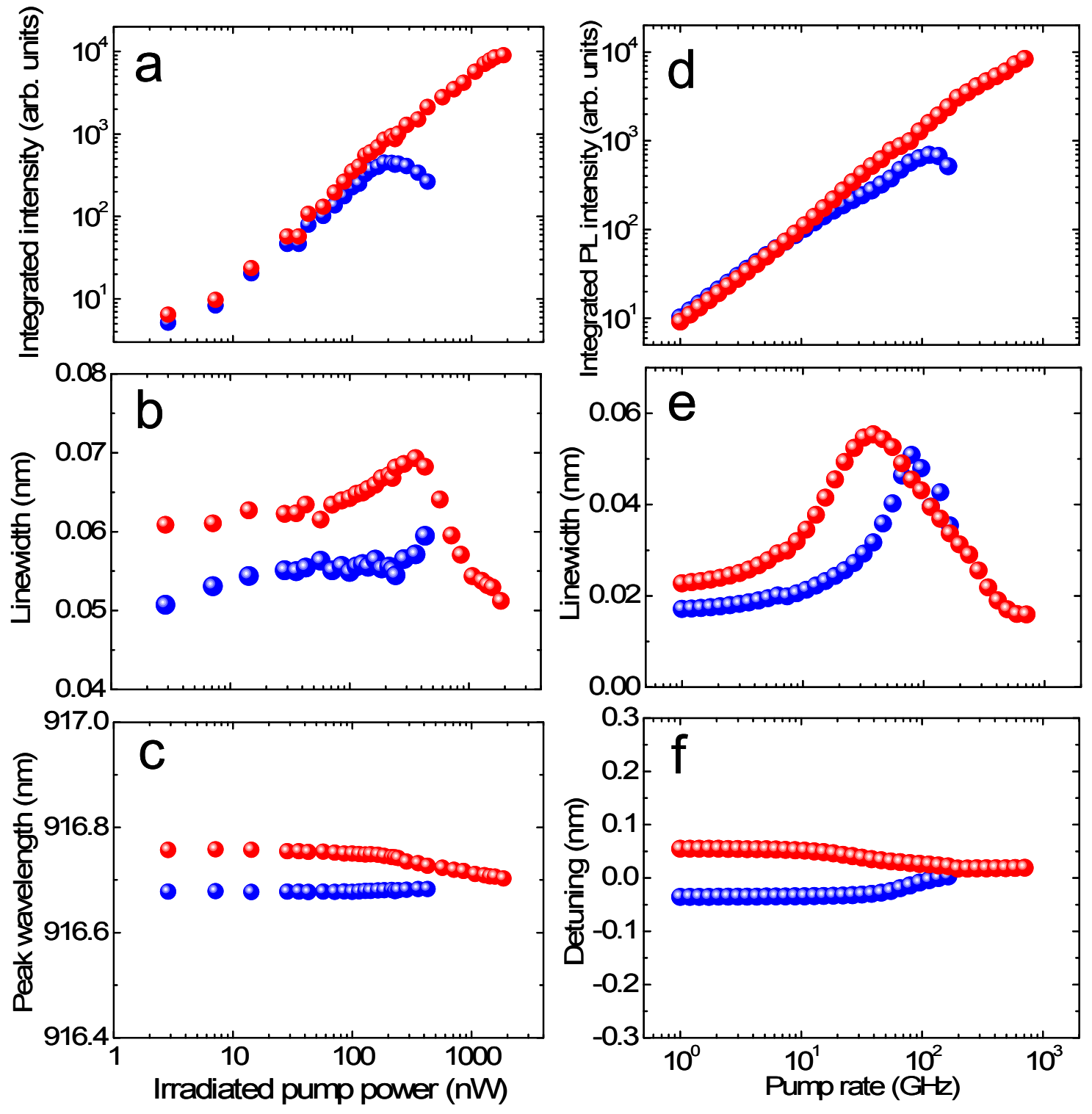

**Figure 3** Analyses of the experimental and computed PL spectra. **a-c.** L-L plots, linewidths, and peak wavelengths obtained by analyzing the measured PL spectra in Fig. 2a. The cavity-like branch (red circle) transits to the lasing regime, while the exciton-like branch (blue circle) attenuated above the threshold by feeding the lasing mode. The peak wavelength of the modes shows a gradual shift from a polariton doublet to a single laser mode at the bare-cavity resonance. **d-f.** L-L plots, linewidths, and peak wavelengths of the computed PL spectra shown in Fig. 2b. The general tendencies are well accorded with those of experimental results.

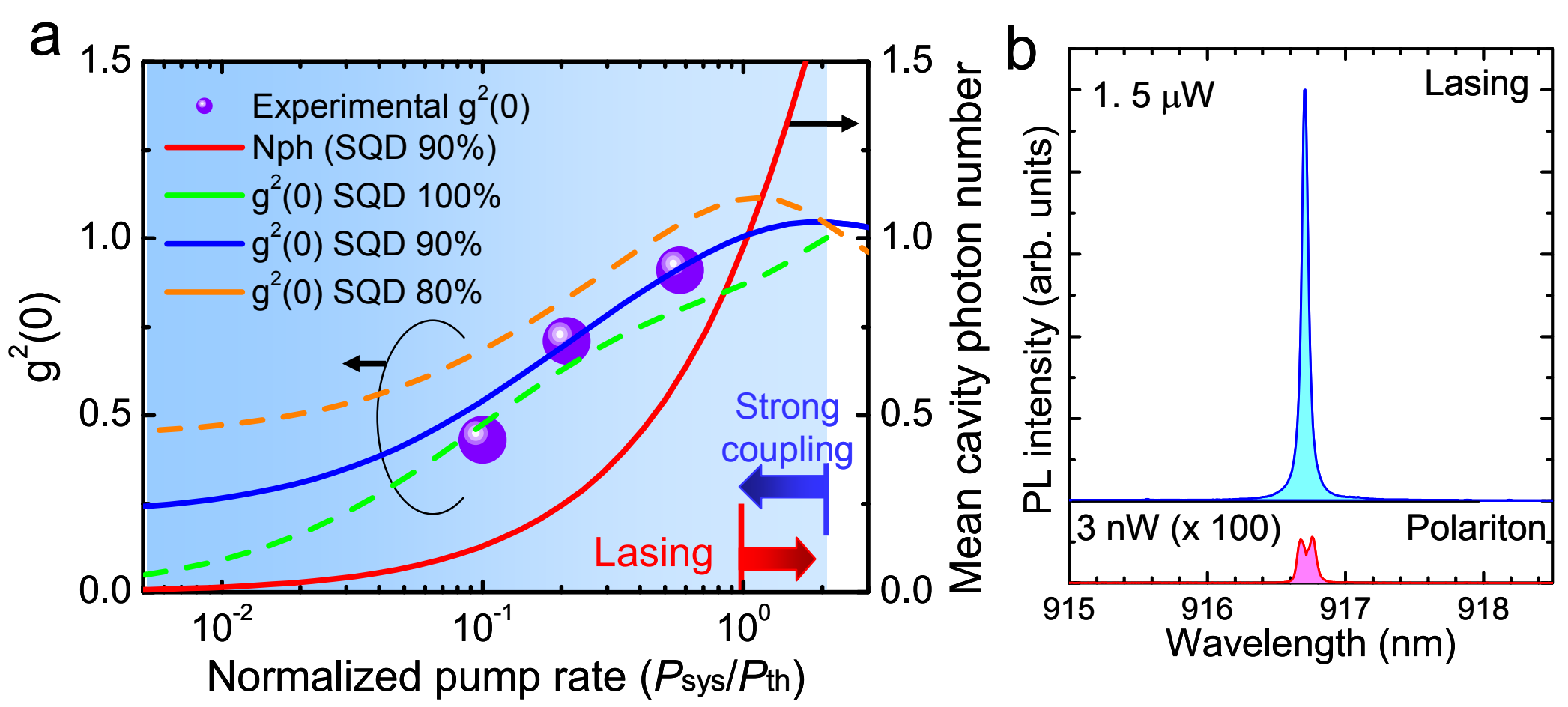

**Figure 4** Mean cavity photon number, $g^2(0)$, and PL spectra in lasing and strong-coupling regimes. **a**, Calculated mean cavity photon number (red line) and $g^2(0)$s with various SQD purities as a function of pump power, and experimental $g^2(0)$ values (purple circle). Laser oscillation begins ($N_{ph} > 1$) in the strong-coupling regime (light-blue region). Comparing the experimental and computed $g^2(0)$ curves reveals that the system has an SQD purity of more than 90% (in the region between the blue and green lines). **b**, PL spectra recorded in lasing (upper panel) and polariton states (lower panel).